# Confinement-Induced Liquid Crystalline Transitions and Chirality Inversion in Amyloid Fibril Cholesteric Tactoids

Gustav Nyström[1], Mario Arcari[1] and Raffaele Mezzenga[1,2]*

[1]ETH Zurich Department of Health Science and Technology, Schmelzbergstrasse 9, LFO E23 Zurich 8092, Switzerland.

[2]ETH Zurich Department of Materials, Wolfgang-Pauli-Strasse 10, • 8093 Zurich, Switzerland.

*Correspondence to: raffaele.mezzenga@hest.ethz.ch

**Chirality is ubiquitous in nature and plays crucial roles in biology, medicine, physics and materials science[1–5]. Understanding and controlling chirality is therefore an important research challenge with broad implications in fundamental and applied sciences. Unlike other classes of chiral colloids, such as nanocellulose or filamentous viruses, amyloid fibrils form nematic phases but appear to miss their twisted form in the phase diagram, the so-called cholesteric or chiral nematic phases[6], and this despite a well-defined chirality at the single fibril level[7]. Here we report the discovery of cholesteric phases in amyloid fibrils, by using β-lactoglobulin fibrils suitably shortened by shear stresses. The physical behavior of this new class of cholesteric materials exhibits unprecedented structural complexity, with confinement-driven ordering transitions between at least three types of nematic and cholesteric tactoids. We use energy functional theory to rationalize these results and demonstrate a chirality inversion upon increasing hierarchical levels, from the left-handed amyloids to the right-handed cholesteric droplets. These findings significantly deepen our understanding of chiral nematic phases and may pave the way to their optimal use in soft nanotechnology, nanomaterials templating and self-assembly.**

The understanding of how chirality is transferred and amplified across various length scales is a fundamental problem with broad implications in nature. Chirality, as observed in nature, typically shows distinct preferences and high selectivity. For instance, only D-sugars and L-amino acids are included in the formation of DNA and proteins, respectively, and this molecular chirality transfers in a way that controls both structure and biological functions. Experimental advances in this area may therefore contribute both to an improved fundamental understanding of the mechanisms behind chirality transfer and also have a direct influence of the design of new biologically-mimicking materials.

Amyloid fibrils are a chiral protein-based system, formed through the self-assembly of β-sheet aggregates into twisted or helical ribbons. Pathological amyloids have great importance in biology[8] and medicine[9], where they are known in relation to neurodegenerative diseases such as Parkinson's and Alzheimer's, while functional amyloids play vital roles for the physical and biological function of living organisms[10–12]. Amyloid fibrils are furthermore emerging as important building blocks for bionanotechnology[7], where they can serve as a versatile platform for new functional biomaterials[13]. Although the amplification, transfer and inversion of chirality is known on the single fibril level for amyloids[14], to date any chiral colloidal liquid crystalline phase, also referred as cholesteric phase, has remained elusive in this system. This is both puzzling and surprising, since their well-defined chirality at the single fibril level, would infer, *a priori,* the presence of cholesteric phases, in direct analogy with other biological systems such as DNA[15], virus,[16] collagen[17] or nanocellulose[18] where chiral nematic phases are routinely observed, yet not fully understood. In stark contrast with other filamentous biological systems, however, cholesteric phases in amyloid fibrils are still to be found. We show in what follows that amyloid fibrils with well-controlled left handedness assemble into cholesteric phases with right handedness. This introduces one additional important system within the class of cholesteric colloids exhibiting handedness inversion, yet opposite to that of previously observed systems where right-handed colloidal particles transfer to left-handed chiral nematic phases[19–21]. We also show that the ensued cholesteric phases present significant complexity in the phase diagram, compared to previously known cholesteric systems and we use scaling concepts on the energy functional to rationalize the essence of the experimental findings, advancing our understanding of cholesteric liquid crystalline systems.

Amyloid fibrils were self-assembled from β-lactoglobulin protein, cut into shorter rod-like particles using shearing and purified through dialysis to remove unreacted protein, see **Fig. 1a**. Transmission Electron Microscopy (TEM) shows a characteristic difference between the semi-flexible nature of the long amyloid fibril and the cut fibrils (Fig. 1b,c), which appear as fully rigid rods. The average length of the cut fibrils, analyzed using contour tracking software[22], is $<L_{avg}>$= 401 nm, n=3500, which combined with the average height from AFM, $<h_{avg}>$= 4 nm[23], gives an aspect ratio of ~100. TEM also confirms the left-handed chirality of the fibrils[7], which is conserved also in the cut and dialyzed fibrils.

When prepared in aqueous dispersions, amyloid fibrils display an isotropic to nematic phase transition that can be well described by the Onsager theory extended to account for the charged and semi-flexible nature of the particles[24–26]. Remarkably, in addition to the nematic phase, the present modified amyloid system displays a rich phase behavior (Fig. 1d–m), where different phases, among which cholesteric droplets, emerge at different equilibration times and sample compositions.

The phase development follows a nucleation and growth behavior where small domains, appearing as bright birefringent droplets under the microscope, nucleate and grow with time until macroscopic phase separation is reached. The exact position within the meta-stable region of the phase diagram determines the fate and the final structure of the observed liquid crystalline droplets. We note that at interfacial tensions as low as those expected here, the trajectories followed by the nucleation and growth of the tactoids can vary significantly from nominal binodal lines, leading to important supersaturation effects and spread in composition of the tactoids. Although most of the observed cholesteric droplets can be classified according to the classification of Bouligand and Livolant[27], three major classes of tactoids dominate the phase diagram: i) nematic tactoids with homogeneous director field, ii) nematic tactoids with bipolar director field, and iii) uniaxial cholesteric droplets.

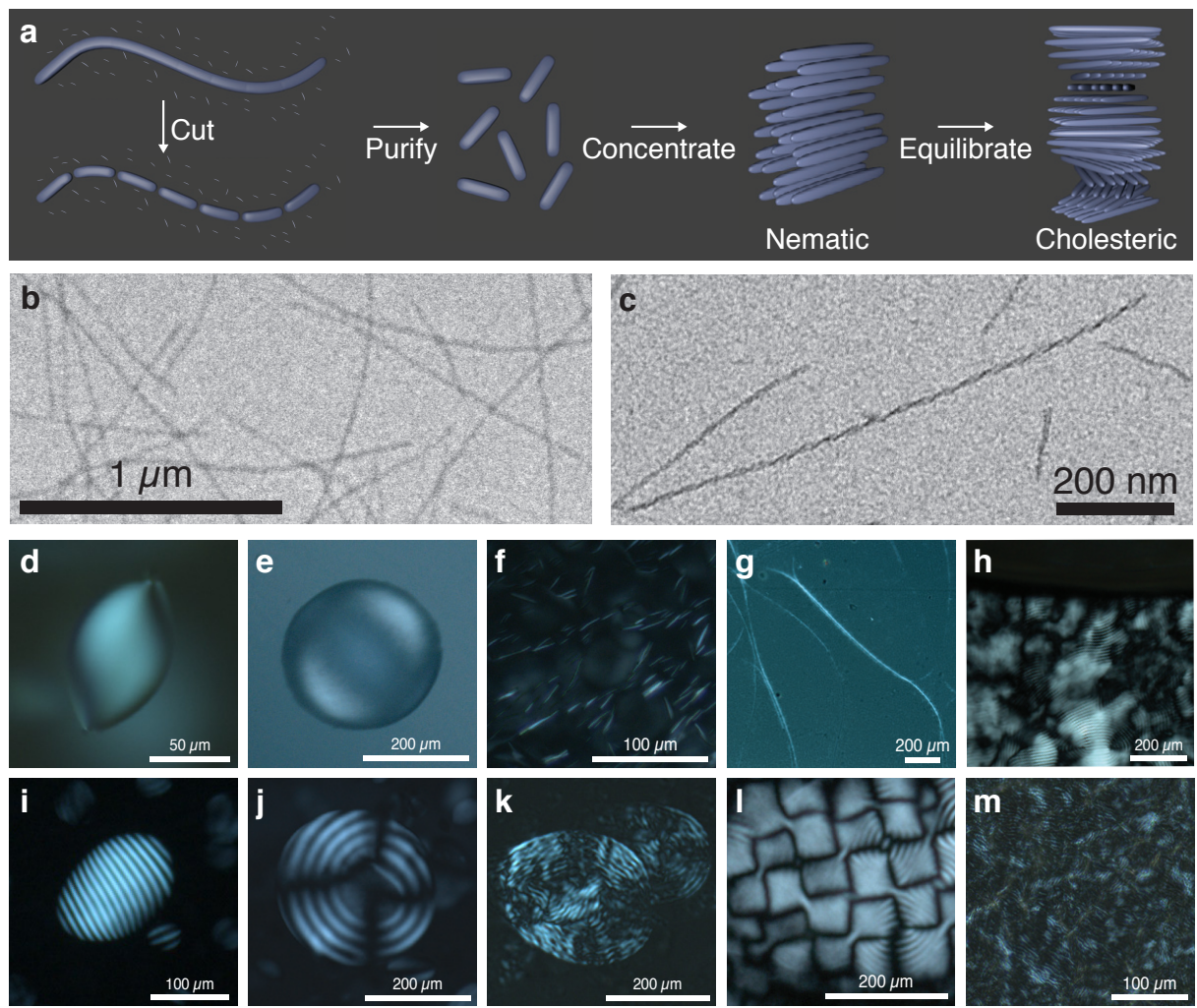

**Fig. 1**. **Amyloid fibril preparation and examples of amyloid fibril liquid crystal polymorphs.** (**a**) Schematic describing the amyloid fibril preparation process and (**b**) TEM images of the real sample. (**d–m**) Polarized optical microscopy images describing: spindle-like nematic tactoids (d), spherical nematic tactoids (e), needle-like tactoids (f); cholesteric tactoids with bands orthogonal to major axis (i), onion-like cholesteric tactoids (j), multi-grain cholesteric tactoids (k); bulk phase boundaries (g), cholesteric focal conics defects (l), isotropic-cholesteric phase boundary (h) and solid film cholesteric phase (m).

To rationalize the most salient features of the observed phase behavior, we rely on the analysis of the liquid crystalline energy landscape, by using a scaling form of the Frank-Oseen elasticity theory. A simplified case has been treated by Prinsen et al.[28] for untwisted, non-chiral droplets, who showed that the energy functional can be reduced into the sum of two terms: the surface contribution, scaling as $\sim r^2$ and an overall elastic contribution, scaling as a volume times the square of a curvature: $r^3 \, r^{-2} \sim r$.

Under these conditions, the theory explains well the transition from homogeneous to bipolar nematic spindle-like droplets when a critical volume is reached: if the droplets are sufficiently small, the elastic term $\sim r$ dominates over the surface $\sim r^2$ term and the elastic energy needs to be minimized primarily by adopting spindle-like shapes of the droplets with a homogeneous nematic field; above a critical volume the energy becomes dominated by the $\sim r^2$ surface term and the energy is now minimized by anchoring the mesogens into a bipolar configuration.

When chirality exists, however, the physics needs to account for the effects arising from the chiral twist. Starting from bipolar spindle-like droplets, as expected for large-enough volumes, the scaling law of the free energy functional can be generalized into:

$$U \sim \frac{Kr^2}{R} + \gamma rR + \frac{1}{2} K_2 q^2 r^2 R \qquad (1)$$

where $r$ and $R = \alpha r$ ($\alpha > 1$) are the short and long radii of the spindle-like droplets; $K=K_1=K_3$ is the elastic constant taken equal to both the splay ($K_1$) and bending ($K_3$) constant, following the one-constant assumption[28], $K_2$ the twist elastic constant, $\gamma$ the interfacial energy and $q$ is the cholesteric wave number, i.e. $2\pi/P$, with P the cholesteric pitch. The first term is the sum of the splay and bend deformation energies, the second term is the surface energy for bipolar orientation of the nematic director and the third term is the elastic contribution due to the chirality.

The question to ask now is: how does a particle of a fixed volume $V \approx r^2 R \approx \alpha r^3$ reshape itself to minimize energy? For very small particles, the first term dominates over the other two terms, provided that $\frac{Kr^2}{R} \gg \gamma R r$ or, taking $R = \alpha r$, $\frac{K}{\gamma \alpha^2} \gg r$, which is equivalent to $\left(\frac{K}{\gamma \alpha^2}\right)^3 \gg \frac{V}{\alpha}$. Thus, for volumes smaller than this threshold, only spindle-like droplets with homogeneous director fields are expected, since the overall energy can be minimized by simply minimizing the first elastic term. For slightly larger droplets, spindle-like droplets assume a bipolar orientation of the nematic director to avoid the anchoring surface term and both first two terms $Kr^4/V + \gamma V/r$ need to be considered to minimize energy: $\frac{\partial U}{\partial r} \sim 4\frac{Kr^3}{V} - \gamma V/r^2 = 0$, which can be reworked into: $R/r = \frac{4K}{\gamma R}$. Thus, for intermediate size spindle-like droplets with bipolar orientation, the aspect ratio decreases with the major size of the droplets[28]. For progressively larger droplets, the $r$ term becomes dominated by the $r^2$ and $r^3$ terms in the energy functional, hence $U \sim \gamma r R + \frac{1}{2} K_2 q^2 r^2 R$, which rewritten as a function of $r$ and $V$ becomes: $U \sim V\gamma/r + \frac{1}{2} K_2 q^2 V$. By minimizing with respect to $r$ and remembering that $V$ is constant and $q$ is a function of $r$:

$$\frac{\partial U}{\partial r} \sim -\frac{V\gamma}{r^2} + K_2 q \dot{q} V = 0 \qquad (2)$$

which is a first order differential equation, whose only physical solution for $q$ is:

$$q = \frac{2\pi}{P} = \frac{\sqrt{q_\infty^2 r - 2\gamma/K_2}}{\sqrt{r}} = \frac{\sqrt{q_\infty^2 (V/\alpha)^{1/3} - 2\gamma/K_2}}{\sqrt{(V/\alpha)^{1/3}}} \qquad (3)$$

where $q_\infty$ is the wave number of the corresponding infinitely relaxed, unconstrained cholesteric phase ($V \to \infty$). This has the important implication that a cholesteric droplet can only be observed for sufficiently large droplets: $\frac{V}{\alpha} > \left(\frac{2\gamma}{q_\infty^2 K_2}\right)^3$. In other words, the larger the ratio between the surface tension and the elasticity constant $K_2$ the larger the droplet must be in order to overcome the chiral symmetry breaking operated by the surface term, as one would expect intuitively. Beyond this critical volume threshold, chiral nematic droplets are observed with a wave number (pitch) asymptotically increasing (decreasing) until reaching the bulk unconstrained $q_\infty$ (pitch $2\pi/q_\infty$) for infinitely large droplets. We also note that eq. 3, which is derived in weak anchoring conditions ($\varpi$ (i·n) << 1), would also predict a critical size for the nematic-cholesteric transition in strong anchoring conditions: in such a case ($\varpi$ (i·n) >>1) and the same conclusions can be drawn by simply replacing $\gamma$ with $\gamma\varpi$ in eq. 3.

In order to test some of these theoretical predictions, an experimental protocol needs to be designed to provide the distinction between homogeneous and bipolar nematic tactoids on one hand, and the measurement of the cholesteric pitch on the other hand. To achieve this we turn to the amyloid fibril water system described in Fig. 1. This system, when prepared at a concentration within the coexistence region of the isotropic-nematic phase diagram, nucleates anisotropic birefringent tactoids that grow with time as shown in **Fig. 2**. When observed under cross-polarized light, nematic tactoids with homogeneous director field show an extinction if the director is oriented parallel to one of the main polarizers axes (Fig 2a), while the curvilinear orientation of the director in the bipolar tactoids leads to a change in texture upon rotation, with a central dark cross, without, however, total extinction (Fig 2b). We note that the presence of this central dark cross is an important feature to identify mirror-symmetric (untwisted) bipolar nematic tactoids, ruling out chiral symmetry breaking operated by the saddle-splay contribution, such as in achiral chromonic liquid crystals[29,30]. This, together with the striped texture of the cholesteric droplets, allows identifying unambiguously the main classes of tactoids found in our system.

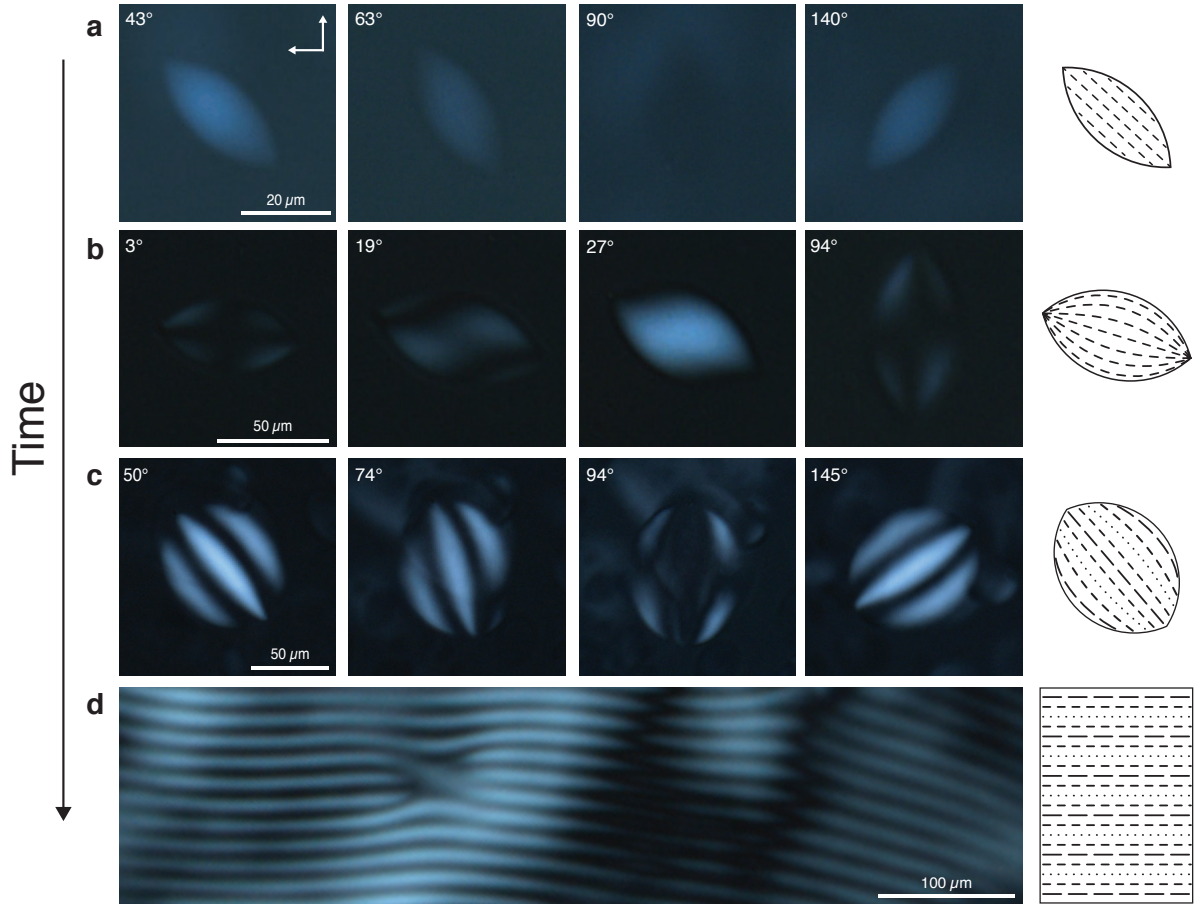

**Fig. 2. Nematic and cholesteric phases of amyloid fibrils as observed by rotating the sample in the plane between fixed crossed polarizers.** (**a**) The smallest nematic tactoids nucleated from the isotropic phase have a homogenous nematic field, leading to extinction upon rotation to an angle parallel with one of the linear polarizers. (**b**) As the tactoids grow with time they convert to a bipolar nematic orientation, with no extinction features. (**c**) With further increase in size the tactoids switch to a cholesteric ordering of the amyloid fibrils. (**d**) In the fully phase separated system a bulk cholesteric phase is formed.

With these experimental data available, it now becomes possible to check the main features predicted by the energy functional. To start, we fit the aspect ratio of bipolar nematic tactoids via $R/r = \frac{4K}{\gamma R}$. As shown in **Fig. 3a** the theory describes well the monotonic decay of the aspect ratio with respect to the major droplet radii. By taking $10^{-6}$ N/m for $\gamma$, typical for water-in-water emulsions[31], the best-fitted value yields $K=4.3 \cdot 10^{-11}$ N, which is a typical order of magnitude for the splay and bend constants of liquid crystals. Secondly, the threshold for the transition of spindle-like droplets from homogeneous to bipolar directors can now also be verified to be $\frac{V}{\alpha} \approx \left(\frac{K}{\gamma \alpha^2}\right)^3 \sim \alpha^{-6} 10^4$ μm³, which, when evaluated using the average observed aspect ratio <α> =1.8, is indeed of the same order of the experimental critical volume below which only spindle-like droplets with homogeneous director are observed (compare with the typical size homogenous tactoid in Fig. 2a).

We found that the tactoids grew either through nucleation and growth of individual tactoids or through coalescence of neighboring tactoids. The time series images in Fig. 3b show two remarkable examples representative of these processes: in the first case (top row), a nematic tactoid grows in size, relatively slowly, becoming a cholesteric tactoid. In the second case (lower row), two nematic tactoids collide forming a single cholesteric droplet. This provides a first direct evidence of the existence of a critical size for the nematic-cholesteric transition, as predicted by the energy functional minimization.

To push the analysis further, the cholesteric pitch was evaluated from the average distance between positions of intensity maxima along a line perpendicular to the cholesteric bands (Fig. 3c) and allowed, combined with the analysis of the tactoid's long and short axis, to quantify the dependence of the cholesteric pitch on the size of the tactoid. By taking, by analogy, the twist constant measured for cholesteric phases of filamentous fd virus in low salt conditions for $K_2$, ~10$^{-12}$ N,[16], 10$^{-6}$ N/m for $\gamma$, and $q_\infty = 2\pi/P_\infty$ where $P_\infty$ is 20 μm, experimentally determined from the cholesteric bulk phase of the amyloid system. This allows plotting the expected evolution of the cholesteric droplet pitch using eq. 3 without using any fitting parameter and is in agreement with the experimental data as shown in Fig. 3d.

Finally, the threshold for nematic-cholesteric droplet transition $\frac{V}{\alpha} \approx \left(\frac{2\gamma}{q_\infty^2 K_2}\right)^3$ can also be determined using the same values as above and is found to be in ~$8.3 \cdot 10^3$ μm³, again, in good agreement with the experimental observations (see Fig. 2c and Fig. 3d).

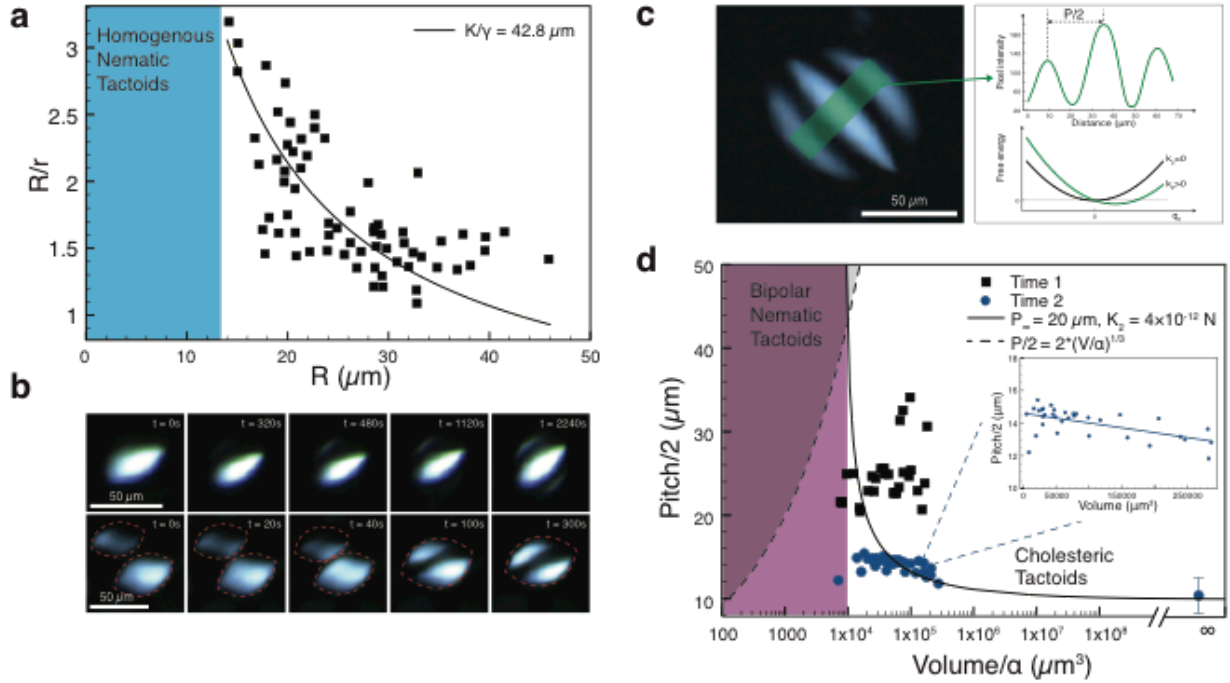

**Fig. 3. Nematic tactoid aspect ratio dependence on tactoid volume, tactoid structural transitions and cholesteric pitch dependence on tactoid volume.** (**a**) The aspect ratio of bipolar nematic tactoids monotonically decreases with increasing tactoid size. (**b**) Growth (top row) and coalescence (bottom row) of nematic tactoids that convert to cholesteric tactoids as their volume increases. (**c**) Schematic of experimental protocol for determination of cholesteric pitch. (**d**) Cholesteric phase volume threshold and evolution of cholesteric pitch as a function of tactoid volume, with experimental values (black square and blue circles) plotted along with the theoretical prediction (black line). As a dashed line we also plot the boundary condition for observing cholesteric droplets, where for P/2 > 2R no cholesteric band can be observed within a tactoid (grey area above dashed line). The inset shows the statistical trend within the blue circles.

**Figure 4a** summarizes the phase diagram emerging from the combined experimental data and theoretical considerations for the observed nematic and cholesteric tactoids in amyloid fibril suspensions. The dispersed phases of amyloid fibrils exhibit a remarkably rich phase diagram of polymorphic liquid crystalline phases, in which order-order transitions are induced by size confinement of the droplets, following the sequences: i) spindle-like tactoids with homogeneous field for $0 < \frac{V}{\alpha} < \left(\frac{K}{\gamma\alpha^2}\right)^3$, ii) spindle-like tactoids with bipolar field for $\left(\frac{K}{\gamma\alpha^2}\right)^3 < \frac{V}{\alpha} < \left(\frac{2\gamma}{q_\infty^2 K_2}\right)^3$, and iii) cholesteric tactoids with asymptotically decreasing pitch for $\frac{V}{\alpha} > \left(\frac{2\gamma}{q_\infty^2 K_2}\right)^3$. The window in the phase diagram for the observation of bipolar nematic droplets, is of the order of $V_{upper}/V_{lower} \approx \left(\frac{2\gamma}{q_\infty^2 K_2}\right)^3 / \left(\frac{K}{\gamma\alpha^2}\right)^3 \approx \left(\frac{2\gamma^2\alpha^2}{q_\infty^2 K K_2}\right)^3 \sim \alpha^6$, and thus is expected to rapidly increase with increasing aspect ratios of the droplets.

With the main features of the phase diagram rationalized through the scaling of the energy functional, we can now return to the question of chirality transfer across length scales. To this end, we designed an experimental setup where the handedness of the cholesteric phase of tactoids could be directly determined. The sample cuvette was oriented at a 45° angle between the crossed polarizers (x- and y-axis in Fig. 4b), so that tactoids with their cholesteric axis parallel to the y'-axis display a maximal cholesteric band intensity. The rotation of the cuvette around the y'-axis is thus also a rotation of the tactoid around its cholesteric axis. This allows the determination of the handedness of the cholesteric phase by observing the rotation induced movement of the cholesteric bands. A right-handed rotation of the cuvette as shown in Fig 4c leads to a downward movement of the maximum intensity bands. This is in agreement with the same rotation and band movement observed for a right-handed helix as schematically shown in Fig 4d allowing us to conclude that the amyloid fibril cholesteric phase is right-handed. As expected, the bands shift by a distance P/4 due to the ~90° rotation. Since the amyloid fibrils have a left-handed chirality, a chirality inversion to right-handedness is observed for the amyloid cholesteric liquid crystalline phase. To the best of our knowledge, this is the first example of a left-handed biological filament generating right-handed cholesteric phases. This widens the known classes of cholesteric phases where chirality inversion from right-handed colloids to left-handed cholesteric phases has been observed[19–21,32]. Different mechanisms have been invoked to explain chirality inversion, such as electrostatics[21], molecular sequence-dependence[32,33], and excluded volume interactions,[34] but no unanimous consensus has been reached yet. Therefore these results may provide additional important experimental evidence towards conclusively solving this paradigm.

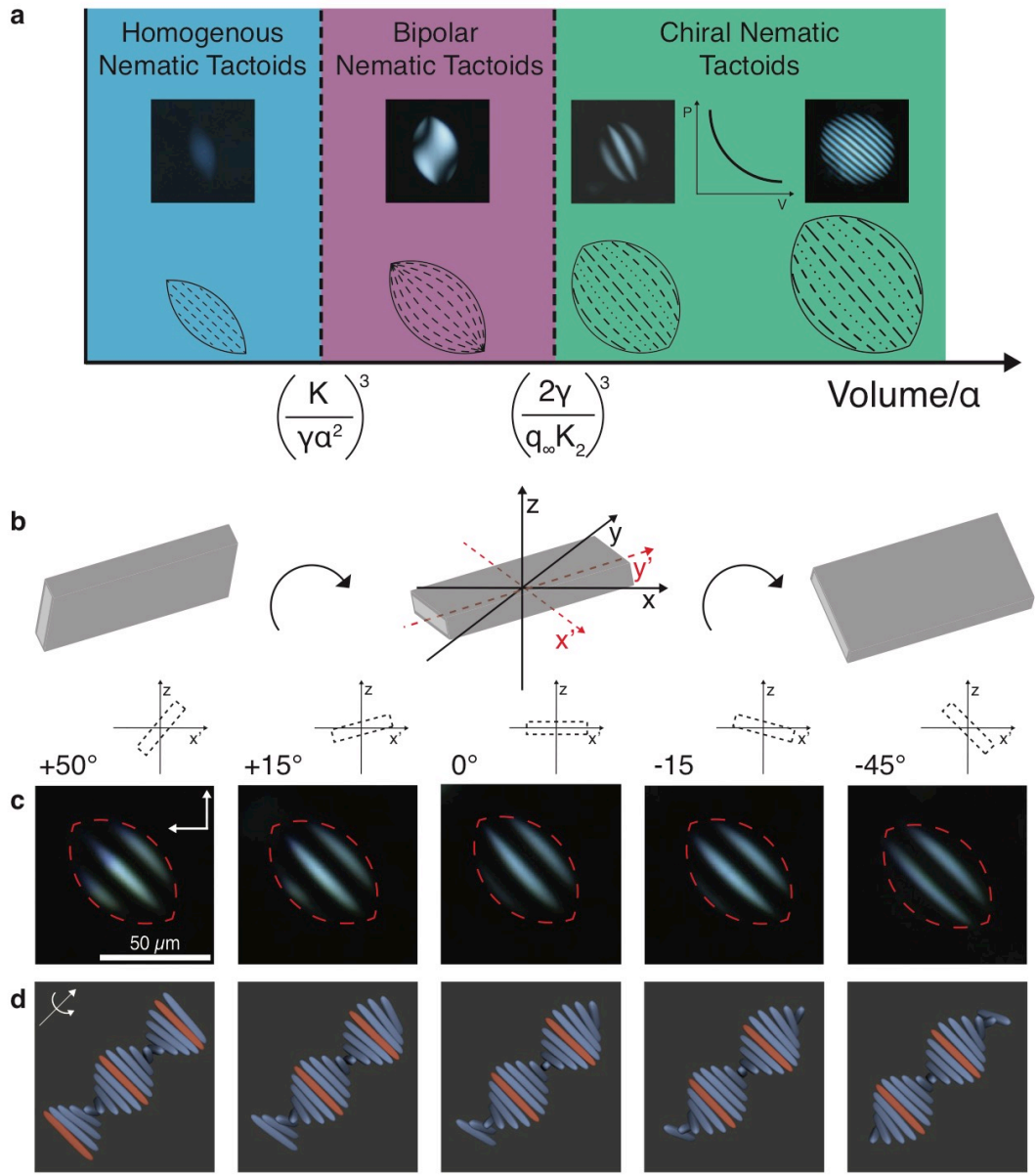

**Fig. 4. Nematic-cholesteric phase diagram and determination of handedness of cholesteric phase.** (**a**) The theory predicts two threshold volumes determining the critical volume for the transition between homogenous to bipolar nematic tactoids and the transition between bipolar nematic tactoids and cholesteric tactoids. (**b**) Rotation of the sample

along its helical axis, oriented at a 45-degree angle to the direction of the polarizer and analyzer, allows a determination of the handedness of the cholesteric phase through the observation of the movement of the cholesteric bands within the tactoid. The downward movement of the cholesteric bands reveals a right handed handedness for amyloid cholesteric droplets (**c**). A right-handed rod helix (**d**) has been added for comparison and the red colored rods illustrate the movement of the maximum intensity birefringence cholesteric band during the rotation.

To conclude, we have discovered the presence of cholesteric phases in amyloid fibril suspensions, revealing a very rich phase diagram in which size-confinement controls order-order transitions among homogeneous nematic, bipolar nematic and chiral nematic tactoids and we have relied on energy functional considerations to rationalize the main experimental findings. We have also demonstrated an inversion of chirality across the increase in hierarchical level from the left-handed amyloids to the right-handed cholesteric. This work introduces a new class of chiral biological nanomaterials and widens significantly the scope and the understanding of cholesteric liquid crystalline phases in general, with possible far-reaching consequences in the design of chiral mesophases, templates and self-assembled materials for bionanotechnology.

**REFERENCES AND NOTES**

1. Salam, A. The Role of Chirality in the Origin of Life. *J. Mol. Evol.* **33,** 105–113 (1991).
2. Ramström, O. & Ansell, R. J. Molecular imprinting technology: Challenges and prospects for the future. *Chirality* **10,** 195–209 (1998).
3. Bradshaw, D., Claridge, J. B., Cussen, E. J., Prior, T. J. & Rosseinsky, M. J. Design, chirality, and flexibility in nanoporous molecule-based materials. *Acc. Chem. Res.* **38,** 273–282 (2005).
4. Shopsowitz, K. E., Qi, H., Hamad, W. Y. & Maclachlan, M. J. Free-standing mesoporous silica films with tunable chiral nematic structures. *Nature* **468,** 422–425 (2010).
5. Gibaud, T. *et al.* Reconfigurable self-assembly through chiral control of interfacial tension. *Nature* **481,** 348–351 (2012).
6. Kitzerow, H. & Bahr, C. *Chirality in Liquid Crystals*. (Springer, 2001).
7. Adamcik, J., Jung, J. & Flakowski, J. Understanding amyloid aggregation by statistical analysis of atomic force microscopy images. *Nat. Nanotechnol.* **5,** 423–428 (2010).
8. Dobson, C. M. Protein folding and misfolding. *Nature* **426,** 884–890 (2002).
9. Glenner, G. G. Amyloid Deposits and Amyloidosis: (First of Two Parts). *N. Engl. J. Med.* **302,** 1283–1292 (1980).
10. Maddelein, M.-L., Dos Reis, S., Duvezin-Caubet, S., Coulary-Salin, B. & Saupe, S. J. Amyloid aggregates of the HET-s prion protein are infectious. *Proc. Natl. Acad. Sci. U. S. A.* **99,** 7402–7407 (2002).

11. Barnhart, M. M. & Chapman, M. R. Curli biogenesis and function. *Annu. Rev. Microbiol.* **60,** 131–147 (2006).
12. Maji, S. K. *et al.* Functional Amyloids As Natural Storage of Peptide Hormones in Pituitary Secretory Granules. *Science.* **325,** 328–332 (2009).
13. Knowles, T. P. J. & Mezzenga, R. Amyloid Fibrils as Building Blocks for Natural and Artificial Functional Materials. *Adv. Mater.* **28,** 6546–6561 (2016).
14. Usov, I., Adamcik, J. & Mezzenga, R. Polymorphism complexity and handedness inversion in serum albumin amyloid fibrils. *ACS Nano* **7,** 10465–10474 (2013).
15. Robinson, C. Liquid-crystalline structures in polypeptide solutions. *Tetrahedron* **13,** 219–234 (1961).
16. Dogic, Z. & Fraden, S. Cholesteric Phase in Virus Suspensions. *Langmuir* **16,** 7820–7824 (2000).
17. Mosser, G., Anglo, A., Helary, C., Bouligand, Y. & Giraud-Guille, M. M. Dense tissue-like collagen matrices formed in cell-free conditions. *Matrix Biol.* **25,** 3–13 (2006).
18. Revol, J., Bradford, H., Giasson, J., Marchessault, R. H. & Gray, D. Helicoidal self-ordering of cellulose microfibrils in aqueous suspension. *Int. J. Biol. Macromol.* **14,** 170–172 (1992).
19. Bonazzi, S. *et al.* Four-stranded aggregates of oligodeoxyguanylates forming lyotropic liquid crystals: a study by circular dichroism, optical microscopy, and x-ray diffraction. *J. Am. Chem. Soc.* **113,** 5809–5816 (1991).
20. Livolant, F. & Maestre, M. F. Circular dichroism microscopy of compact forms of DNA and chromatin in vivo and in vitro: cholesteric liquid-crystalline phases of DNA and single dinoflagellate nuclei. *Biochemistry* **27,** 3056–3068 (1988).
21. Tombolato, F., Ferrarini, A. & Grelet, E. Chiral Nematic Phase of Suspensions of Rodlike Viruses: Left-Handed Phase Helicity from a Right-Handed Molecular Helix. *Phys. Rev. Lett.* **96,** 258302 (2006).
22. Usov, I. & Mezzenga, R. FiberApp: An Open-Source Software for Tracking and Analyzing Polymers, Filaments, Biomacromolecules, and Fibrous Objects. *Macromolecules* **48,** 1269–1280 (2015).
23. Lara, C., Adamcik, J., Jordens, S. & Mezzenga, R. General self-assembly mechanism converting hydrolyzed globular proteins into giant multistranded amyloid ribbons. *Biomacromolecules* **12,** 1868–1875 (2011).
24. Mezzenga, R., Jung, J.-M. & Adamcik, J. Effects of charge double layer and colloidal aggregation on the isotropic-nematic transition of protein fibers in water. *Langmuir* **26,** 10401–10405 (2010).
25. Vroege, G. J. & Lekkerkerker, H. N. W. Phase transitions in lyotropic colloidal and polymer liquid crystals. *Reports Prog. Phys.* **55,** 1241–1309 (1992).
26. Khokhlov, A. & Semenov, A. Liquid-crystalline ordering in the solution of long persistent chains. *Physica A* **108,** 546–556 (1981).
27. Bouligand, Y. & Livolant, F. The organization of cholesteric spherulites. *J. Phys.* **45,** 1899–1923 (1984).
28. Prinsen, P. & van der Schoot, P. Shape and director-field transformation of tactoids. *Phys. Rev. E* **68,** 21701 (2003).
29. Tortora, L. & Lavrentovich, O. D. Chiral symmetry breaking by spatial confinement in tactoidal droplets of lyotropic chromonic liquid crystals. *Proc. Natl. Acad. Sci. U. S. A.* **108,** 5163–5168 (2011).
30. Nayani, K. *et al.* Spontaneous emergence of chirality in achiral lyotropic chromonic liquid crystals confined to cylinders. *Nat. Commun.* **6,** 8067 (2015).

31. Balakrishnan, G., Nicolai, T., Benyahia, L. & Durand, D. Particles trapped at the droplet interface in water-in-water emulsions. *Langmuir* **28,** 5921–5926 (2012).
32. Zanchetta, G. *et al.* Right-handed double-helix ultrashort DNA yields chiral nematic phases with both right- and left-handed director twist. *Proc. Natl. Acad. Sci. U. S. A*. **107,** 17497–17502 (2010).
33. Frezza, E., Tombolato, F. & Ferrarini, A. Right- and left-handed liquid crystal assemblies of oligonucleotides: phase chirality as a reporter of a change in non-chiral interactions? *Soft Matter* **7,** 9291 (2011).
34. Belli, S., Dussi, S., Dijkstra, M. & van Roij, R. Density functional theory for chiral nematic liquid crystals. *Phys. Rev. E* **90,** 20503 (2014).

**Acknowledgements** Stephan Handschin is acknowledged for help with the sample rotation stage for polarized optical microscopy. Salvatore Assenza and Cristiano de Michele are acknowledged for valuable discussions.

**Author Contributions** GN designed and carried out the experiments, analyzed data and interpreted the results. MA carried out the experiments, analyzed data and interpreted the results. RM developed the theoretical description, analyzed and interpreted the results and designed the study. All authors discussed the results and contributed to writing.

**Additional Information** Correspondence and requests for materials should be addressed to RM (raffaele.mezzenga@hest.ethz.ch).

**Competing Financial Interest** Authors declare no competing financial interest.